# Deciphering the Effect of Traps on Electronic Charge Transport Properties of Methylammonium Lead Tribromide

Artem Musiienko[1*], Jindřich Pipek[1], Petr Praus[1], Mykola Brynza[1], Eduard Belas[1], Bogdan Dryzhakov[2], Mao-Hua Du[3], Mahshid Ahmadi[2**], Roman Grill[1]

[1]*Charles University, Faculty of Mathematics and Physics, Institute of Physics, Ke Karlovu 5, CZ-121 16, Prague 2, Czech Republic*

[2]*Joint Institute for Advanced Materials, Department of Materials Science and Engineering, University of Tennessee, Knoxville, TN 37996, USA*

[3]*Oak Ridge National Laboratory, Materials Science and Technology Division, Oak Ridge, USA*

Corresponding authors emails:

[*]musiienko@karlov.mff.cuni.cz,

[**]mahmadi3@utk.edu

**Abstract:**

Organometallic halide perovskites (OMHPs) have undergone remarkable developments as highly efficient optoelectronic materials for a variety of applications. Several studies indicated the critical role of defects on the performance of OMHP devices. Yet, the parameters of defects and their interplay with free charge carriers remain unclear. In this study we explore the dynamics of free holes in methylammonium lead tribromide ($MAPbBr_3$) single crystals using the time of flight (ToF) current spectroscopy. By combining the current waveform (CWF) ToF spectroscopy and the Monte Carlo (MC) simulation, three energy states were detected in the band gap of $MAPbBr_3$. Additionally, we found the trapping and detrapping rates of free holes ranging from a few μs to hundreds of μs and, contrary to previous studies, a strong detrapping activity was revealed. It was shown that these traps have a significant impact on the transport properties of $MAPbBr_3$ single crystal devices, including drift mobility and mobility-lifetime product. To demonstrate the impact of traps on the delay of free carriers, we developed a new model of the effective mobility valid for the case of multiple traps in a semiconductor. Our results provide a new insight on charge transport properties of OMHP semiconductors, which is required for further development of this class of optoelectronic devices.

**Introduction**

Despite the impressive progress in the commercialization of Si photovoltaics (PVs) in the last decade, the Si-based single-junction solar cell efficiency is ultimately limited by the theoretical limit of 29%[1] and the cost needs to be further reduced to compete with fossil fuels. Therefore, the current research aims at searching for alternative materials with high efficiencies and low costs and improving tandem solar cell designs. Recently, the inexpensive organometallic halide perovskite (OMHP) semiconductors have emerged as a new class of PV materials with highly efficient light absorption and charge transport The efficiency of the OMHP solar cells increased significantly from 3.8% in 2009 to 23.7%[2,3] in 2019. Another great advantage of OMHP semiconductors is the fabrication capability on flexible substrates[4–7], which offers an additional opportunity for the development of portable power sources[8] and new PV architectures[9,10]. Among the wide compositional range of OMHPs, methylammonium lead tribromide perovskite ($MAPbBr_3$) has attracted a great interest for its potential applications in tandem solar cells[11–16], as well as in high energy radiation sensors[17–20], photodetectors[21,22], light-emitting diodes[23,24], and lasers[21].

One of the most critical factors in the performance of multifunctional OMHP devices is the presence of trapping centers resulting in the loss of charge collection efficiency in a solar cell or a detector. Trapping centers in a semiconductor lattice form energy states in the band gap[25–28]. These energy states affect the relaxation dynamics of free carriers by trapping and therefore, detrimentally influence the free charge transport properties such as lifetime and drift mobility. The detection and characterization of such traps and their associated relaxation dynamics are highly challenging. The dominant non-radiative nature of such energy transitions does not allow the measurement of key parameters related to trapping/detrapping by optical spectroscopies including photoluminescence[29–31] because the optical and thermal transition energies of traps in semiconductors are different.[32–34] In addition, optical spectroscopies cannot detect shallow traps with energies $E_t < 0.3$ eV due to the strong Urbach tail absorption[35,36]. Other techniques such as thermal emission can be limited by the low activation energy of shallow traps. The presence of several phase transitions in OMHPs also prevents adequate cooling of the sample to reveal the properties of traps via thermal relaxation of traps[37–39].

Recently, by combining the current waveform Time of Flight (CW ToF) and the photo-Hall effect spectroscopy (PHES) we revealed deep levels and their relative positions in the bandgap of $MAPbBr_3$ single-crystal devices[27]. Several studies observed similar deep energy transitions by optical excitation methods[40,41]. In addition to deep levels, the presence of multiple shallow levels in OMHPs have been estimated theoretically[42–47]. Yet, the trapping parameters – trapping and detrapping time constants– of these levels and the interplay of free charge carriers with these energy levels have not been shown experimentally.

The primary aim of this study is to uncover the effect of traps on charge transport dynamics in MAPbBr$_3$ single crystal devices using the ToF current spectroscopy. The ToF current spectroscopy is based on analyzing a transient photocurrent generated by charge carriers drifting through the sample under an applied bias. Generally, traps in a material affect the charge cloud and as a result, the measured current waveforms (CWFs). Thus, CWF provides valuable information on free carrier relaxation including trapping and detrapping times from the trap states in the bandgap. A schematic of the working principle of this method is shown in **Figure 1**. More information of the technique can be found in methods, supplementary information-S1and S2 and in our previous studies[27,48,49].

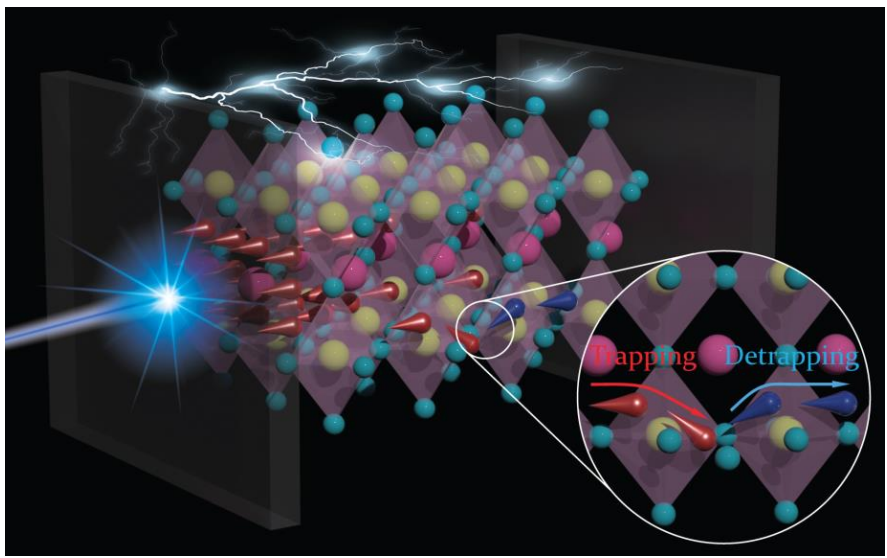

**Figure 1.** A schematic of physical principle of the ToF method. The light pulse generates free carriers (holes) near the electrode (anode). The photogenerated free holes further drift towards the cathode by an electric pulse bias. The drifting holes interact with traps and their interaction affects the current transient.

The analytical calculation of ToF CWFs (based on the current continuity and drift-diffusion equation) are limited to only simple examples (e.g. a solution of the drift-diffusion equation with a single trap)[50–53]. Thus, numerical simulations are necessary in case of multiple trapping states.

Monte Carlo (MC) simulation is known as a powerful and convenient method for studying charge dynamics. To model the experimental results of ToF spectroscopy and to decipher the effect of trapping and detrapping of carriers from defect states, we first solve charge transport equations by one dimensional (1D) MC simulation[54]. In this simulation, each MC particle represents either a free hole drifting with a velocity of $\mu_h E(x)$ in the valence band (where $\mu_h$ is the drift mobility and $E(x)$ is the applied electric field) or a trapped hole in one of the states in the band gap. Here, using the MC simulation we develop a charge transport model, including non-radiative energy transitions associated with traps. We simulate how the trapping and detrapping of photogenerated charge carriers limit the drift of free carriers in a MAPbBr$_3$ single crystal device at different electric biases relevant to the device operation. In this approach, the simulated trapping and detrapping of charge carriers provide a novel insight into the free carrier dynamics and charge transport across the bulk of a MAPbBr$_3$ single crystal device. The proposed model is further validated by the ToF measurement.

It is known that the predicted multiple trapping states in the band gap of OMHP complicate the dynamics of charge transport beyond the classical model of trap controlled mobility[55,56]. Such a model, also known as the effective mobility model, considers a semiconductor with a single trap delaying free charge carriers. Therefore, a new approach is necessary to unambiguously describe the dynamics of free charge carrier transport in OMHP semiconductors. To do this, we use MC simulations to investigate the delay of charge carriers and identify the effective transit time by tracking the center of the charge cloud affected by traps. We then reassess the definition of the effective mobility ($\mu_h^{eff}$) according to the effective transit time. Such an approach allows us to study charge transport properties in any semiconductor with any number of active traps. Finally, we analyze the effect of traps on the effective mobility of carriers (holes) and its relationship with the electric field, thickness, and temperature in OMHP devices.

In addition, to demonstrate the effect of traps on the performance of MAPbBr$_3$ single crystal devices we explore the influence of traps on the mobility-lifetime product. Understanding the impact of traps on charge transport properties is necessary to control the traps and to further improve the performance of OMHP devices.

**Results and discussion**

**Charge Transport Dynamics in Single crystals of MAPbBr$_3$ perovskite**

The temporal dynamics of free carriers is considered as the most crucial characteristic of a material, defining the efficiency of a semiconductor device. The temporal dynamics, given by Equation (1), is mainly affected by energy states in the band gap.

$$\frac{dp}{dt} = \sum_i \left( -\frac{p}{\tau_{Ti}} + \frac{p_{ti}}{\tau_{Di}} \right) \tag{1}$$

Here, the effect of traps on the free hole concentration ($p$) is described by the specific trapping ($\tau_{Ti}$) and detrapping times ($\tau_{Di}$) of the *i*-th level; and $p_{ti}$ is the concentration of holes trapped at the *i*-th level. Depending on the trapping and detrapping times, defects can induce short-term and long-term trapping of free carriers. The presence of traps that cause fast trapping and detrapping in a material delays the free carrier drift and consequently reduces the drift mobility. Theoretically, the long-term trapping is commonly associated with carrier lifetime ($\tau_{life}$). However, in reality, a trapping center releases the trapped carriers after $\tau_{Di}$, and the detrapped free carriers continue moving through a semiconductor. This effect, which is missing in the conventional carrier lifetime[57] and effective mobility models, leads to errors in the study of transport features and charge collection properties in semiconductor devices.

By probing ToF signals in MAPbBr$_3$ single crystal devices we found reliable hole signals in the CWFs, but the electron signals could not be revealed (**Figure 2**a). Therefore, here we only study the free hole transport. By studying the transit time ($T_R$) of the ToF CWFs, i.e. the time required for the holes to transit through the semiconductor, and the relaxation dynamics before and after $T_R$, we can reveal the information of traps that interact with holes.

Figure 2 shows the ToF transient current from holes collected by a 100 V electric bias. The 20 nm semi-transparent Cr electrode (anode) immediately collects the electron cloud, and only free holes drifting through the bulk of MAPbBr$_3$ towards the cathode induce ToF signals. As shown in Figure 2a the profile of the hole drift in MAPbBr$_3$ reveals a complex relaxation dynamics: two exponentially-decaying regions before the transit time ($T_R$=32 µs), and a long current tail after $T_R$. The distinct transit time region indicates an insignificant impact of the free hole recombination as a fraction of the charge cloud reached the cathode and produced a transit time bending[58] (change of the *I*(*t*) curvature) of CWF. The long $T_R$ indicates a long lifetime of holes ($\tau_{life} > T_R$) and a relatively low free hole mobility in the MAPbBr$_3$ single crystal device.

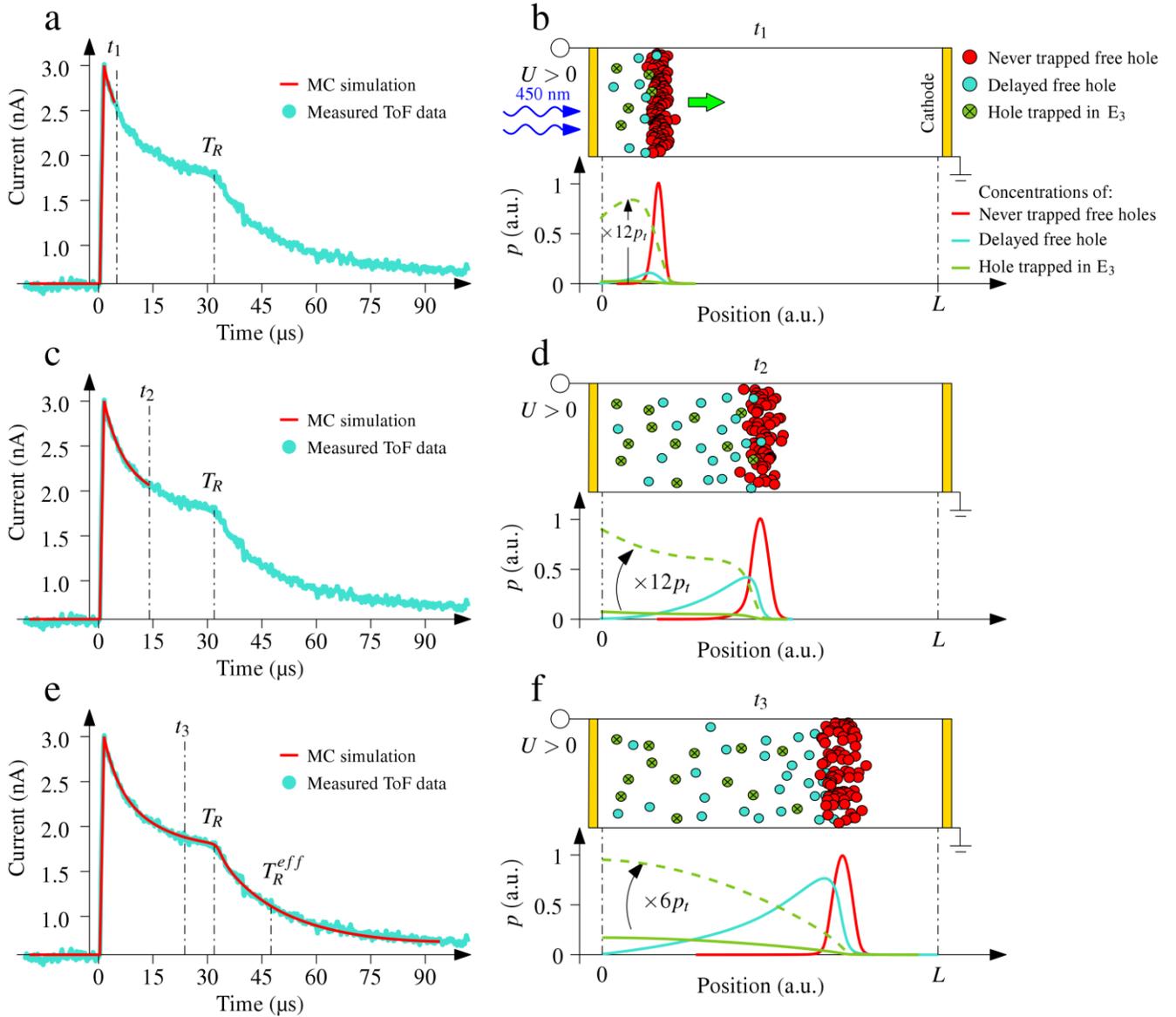

**Figure 2.** a), c), and e) ToF CWF measured at 100 V. The red curve shows the best MC simulation fit with three energy levels at different times $t_1 = 6$, $t_2 = 14$, and $t_3 = 23$ μs respectively. b), d), and f) the top panels demonstrate the time evolution of the free charge cloud (subdivided according to their trapping-detrapping history) during the drift process at $t_1$, $t_2$, and $t_3$, respectively. The bottom panels show the respective normalized concentrations of never trapped holes, trapped and delayed holes in the material. The attached video (**video 1** in supplementary materials) shows the evolution of the charge cloud computed by the MC simulation with a 0.1-μs time resolution, which allows the direct visualization of the free charge movement in a MAPbBr$_3$ device.

To explore the complex charge dynamics and to assess transport parameters of MAPbBr$_3$ we applied a MC simulation in combination with the least square regression analysis. The details of the MC simulation (Figure S3 and Equations S1-S6) and the fitting method are shown in supplementary materials (Figure S5 and Table S1). The red line in Figure 2a, c, and e represent the MC transport model with minimum deviations from the ToF spectra. This MC model with parameters summarized in Table 1 reveals that three trap states affect the hole transport. According to the simulated MC transport model (**Figure 3**), the two energy levels $E_1$ and $E_2$ cause the fast trapping of free holes with nearly the same trapping time of 23 and 24 µs. These levels shortly release the trapped holes to the valence band with detrapping times of 3 and 14 µs, respectively. The trap level, $E_3$, gives rise to the long-lasting carrier trapping with a trapping time of 90 µs. This energy level shows a relatively slow ($\tau_{D3} > T_R$) detrapping time of 120 µs.

**Table 1** Charge transport parameters of single-crystal MAPbBr$_3$ found from the combination of ToF and MC simulation

| Energy level | Trapping time (µs) | Detrapping time (µs) | Trapping/detrapping ratio | $\sigma_h \times n_t$ product* $10^{-3}$ cm$^{-1}$ |
|:---:|:---:|:---:|:---:|:---:|
| $E_1$ | 23 | 3 | 7.7 | 1.31 |
| $E_2$ | 24 | 14 | 1.7 | 1.25 |
| $E_3$ | 90 | 120 | 0.75 | 0.34 |

*Hole capture cross section and trap density product.

To study the effect of traps on charge transport, it is convenient to separate the free carrier profiles from carrier trapping and detrapping phenomena. Using the MC simulation, we divide the free holes into three groups: never trapped holes (holes which did not interact with any traps), delayed holes (holes detrapped at least once by any traps), and long-term trapped holes (holes trapped by the level $E_3$). Figures 2b, d, and f represent the simulated evolution of a free-hole cloud and its interaction with traps during the drift through a MAPbBr$_3$ single crystal device under the electric bias of 100 V at different times ($t_1 = 6$ µs, $t_2 = 14$ µs, and $t_3 = 23$ µs). Clearly, the $E_1$ and $E_2$ traps decelerate free holes by relatively fast trapping-detrapping processes and, as a result, the fraction of delayed holes incraeses with time. As can be seen in the bottom panel of Figure 2f, when the charge cloud reaches the collecting electrode, the concentration of delayed holes is comparable with the concentration of never trapped holes. Thus, the detrapped carriers create an extended profile of delayed holes, which deviate from the total charge cloud in the Gaussian distribution (see blue and red lines in Figure 2f).

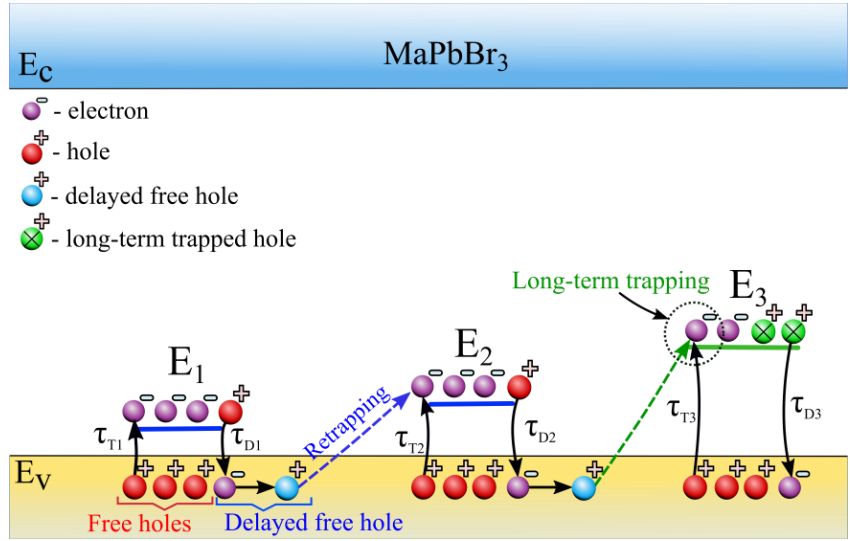

**Figure 3.** Charge transport model in MAPbBr$_3$. Upward and downward arrows illustrate energy transitions (trapping and detrapping). Table 1 summarizes the parameters of these transitions estimated by ToF measurements and MC simulations.

In contrary to the fast trapping-detrapping dynamics, the trap E$_3$ causes the long-term trapping of holes which remain trapped in this defect ($p_t$) during the whole drifting process (see Figure 2f). The average delaying of the hole cloud by traps or the effective transit time was estimated from the MC simulation to be 48 μs (see Figure 2e), while $T_R$ = 32 μs reflects the transit time of the never trapped holes.

Here, we revealed the presence of traps, their parameters, and their roles in delaying free hole transport in a MAPbBr$_3$ single crystal device under the bias of 100 V. It is expected that the levels E$_1$, E$_2$, and E$_3$ are located near the valence band as they have a direct influence on free hole transport in MAPbBr$_3$.

**Validation of the charge transport model and uniform electric field profile**

It is well established that the drift velocity ($v_{dr}$) of charge carriers under a uniform electric field ($E$) is directly proportional to the electric field and the free carrier mobility. Therefore, free carriers driven by a lower bias need longer time to be collected by the electrode (for details see Figure S4b and Equation S7 in supplementary materials). To validate the simulated MC transport model and to study the effect of electric field on the transit time and hole trapping dynamics, CWF ToF measurements were performed at different biases (**Figure 4**a). The results of MC simulations (based on parameters in Table 1) agree very well with the experimentally measured bias-dependent CWFs and follow the main trends of ToF results, including the sharp current decrease at the beginning of CWF and the long tail after $T_R$. The effect of the trap $E_3$ is even more evident at lower biases between 20 to 80 V. As can be seen in Figure 4b, both experimental CWFs and MC simulations follow the same trend of a single trap with slow detrapping (level $E_3$) in the charge transit region, $t < T_R$. The good agreement between the simulated and measured results at all biases further confirms the reliability of the MC simulation and supports the proposed explanation of charge transport dynamics.

In addition to short and long-term charge trapping, defects can induce an electric field distortion by creating a depletion region near the metallic electrode[48]. Several studies demonstrated the presence of mobile defects in OMHPs and discussed that the collection of these low-mobility species at the interface could lead to the deformation of the electric field profile[59]. To suppress the possible formation of the space charge in the sample during a ToF experiment, we use a short voltage pulse of 1 ms, synchronized with a light pulse of 100 ns (see details in supplementary materials and method section). It should be noted that the drift of photo-induced carriers across the material can also result in an electric field distortion. Here, by integrating the CWF in Figure 4a, a low carrier concentration of ~$10^6$ cm$^{-3}$ and an electric field distortion of 0.6 Vcm$^{-1}$ are obtained (see details in supplementary materials Equation S8). Therefore, the effect of free charge carriers on the electric field distortion is negligible.

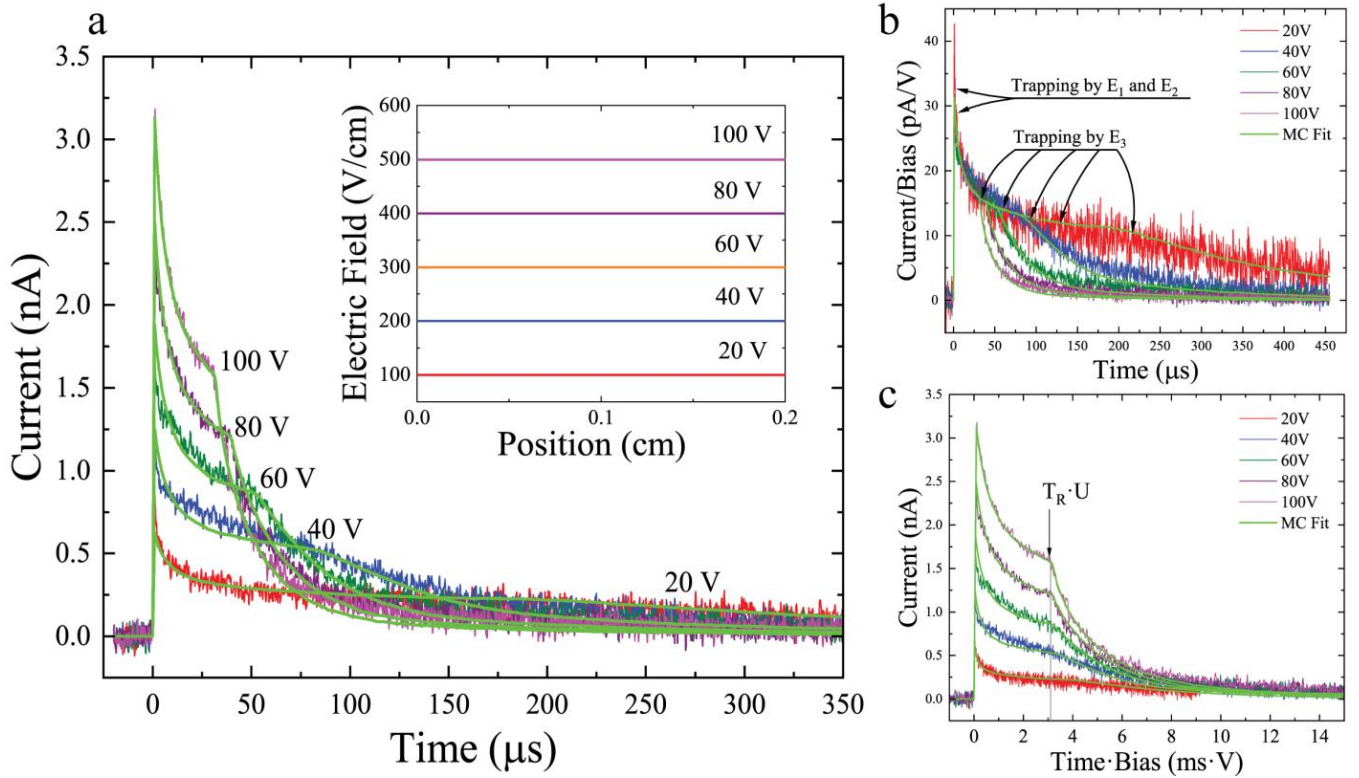

**Figure 4.** a) Bias dependence of CWFs. The green curves represent the simulated MC fit. The inset shows electric field profiles between the two electrodes. b) Normalized CWFs at different biases. c) CWFs dependence on normalized time according to Equation 2.

CWFs at different biases can be used to verify the electric field profile and the presence of the space charges. The presence of non-negligible space charges deforms the electric field. The deformed electric field systematically prolongs the transit time of the never-trapped charge cloud at different biases[58]. Thus, the transit times in ToF CWFs follow Equation 2 in a semiconductor with space charges:

$$T_R(U) \cdot U \neq const \quad (2)$$

According to CWFs ToF and MC simulations in Figure 4c, all CWFs have the same $T_R(U) \cdot U$ product, which confirms the uniform distribution of the electric field in MAPbBr$_3$ single crystal devices.

**Charge distribution in MAPbBr$_3$ induced by trap states at different biases**

In the previous section it is shown that the fast and slow trapping-detrapping processes induced by three energy states in the band gap affect the charge dynamics in MAPbBr$_3$ devices. To further understand the impact of each energy level on the charge transport at different electric biases (100, 20, and 1 V), the time evolution of the charge cloud is studied using both the MC simulation and ToF CWF (**Figure 5**). Interestingly, we found that the relaxation dynamics of free holes changes with bias (Figure 5a, c, and e) and the number of trapping-detrapping events increases at lower electric bias. These changes can be attributed to the interplay between traps and free holes. As can be seen in Figure 5b, d, and f, the lower the bias, the higher the concentration of delayed holes is produced in MAPbBr$_3$. This is because the charge cloud needs more time to drift across the material and, as a result, there is a higher possibility to interact with traps.

It is proposed that the interplay between free holes and traps follows four regions as illustrated in Figure 5a, c, d. After the free holes are generated by a light pulse, they start to drift through the bulk. All traps actively capture the free holes, leading to the occupation of all traps and, consequently, the reduced concentration of free holes in the region **(i)**. In this region, the charge trapping induced by the traps E$_1$ and E$_2$ dominates with faster trapping times. The trap E$_1$ first reaches a saturation point (a steady-state condition, in which the trapping and detrapping rates are equal) due to the faster detrapping time from this trap ($\tau_{D1} > \tau_{D2} > \tau_{D3}$). Next, the holes detrapped from E$_1$ are retrapped by the traps E$_2$ and E$_3$. Usually, in a material with a single trap, the occupation of a trap does not change after a steady state condition is reached. However, here due to the presence of three traps the occupation of the trap E$_1$ starts to decrease after the saturation due to the trapping by other traps (E$_2$ and E$_3$). We note that the cross retrapping process of the delayed holes can play an important role, further delaying the charge cloud.

Similar to the trap E$_1$, the trap E$_2$ reaches a steady state condition at region **(ii)**. When the traps E$_1$ and E$_2$ both attain their steady-state conditions, they do not capture additional free holes; therefore, the relaxation dynamics. Figures 5b), d), and f) show the simulated visualization of the spatial distribution of traps (top panel), the distribution of occupied trap density (middle panel), and the profile of the charge cloud in MAPbBr$_3$ at different biases (bottom panel). The attached videos (videos 2-4) show the simulated evolution of the charge cloud and trap occupation with 0.1 µs time resolution allowing direct visualization of the interplay between free holes and a particular trap.

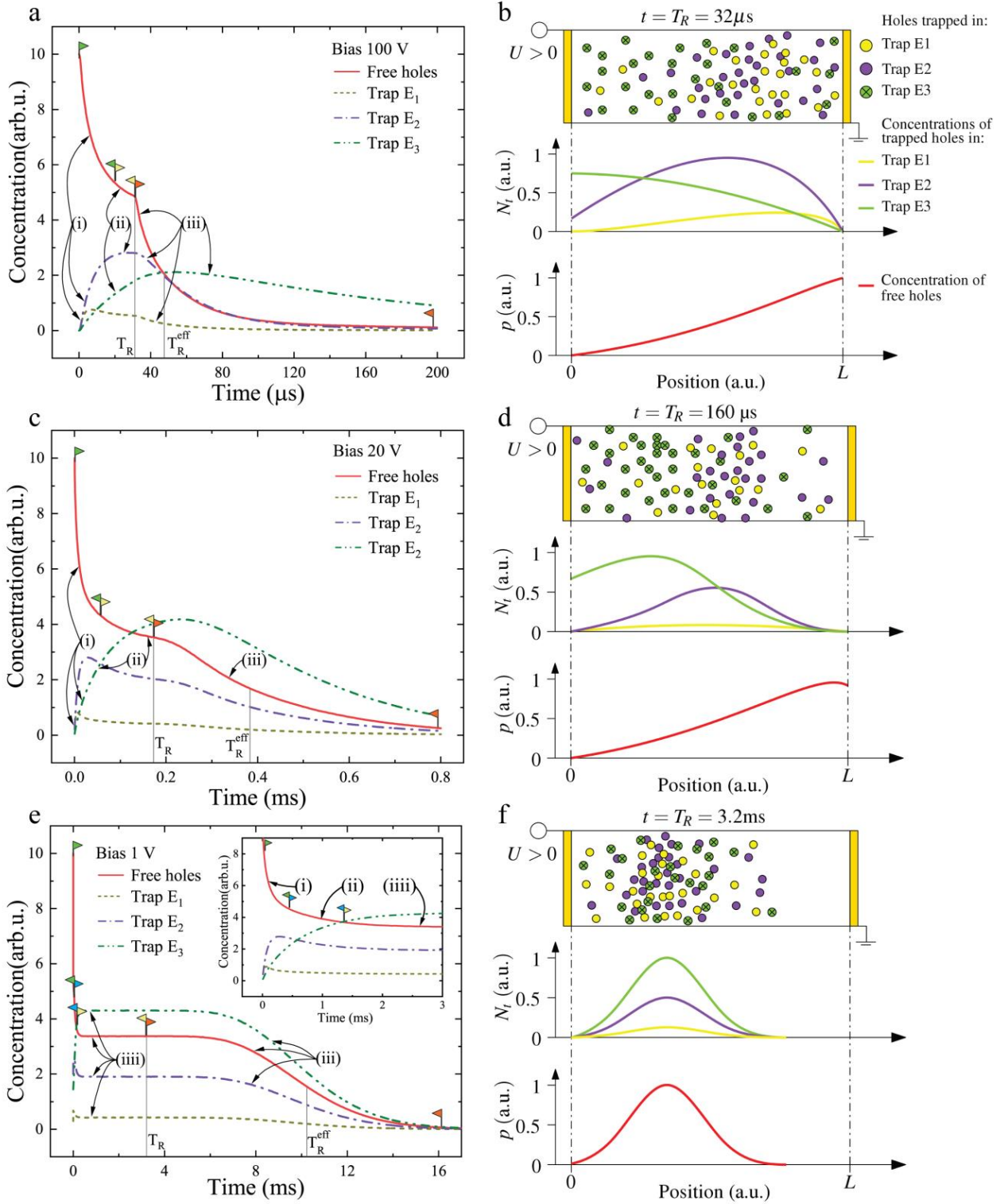

**Figure 5.** The evolution of free holes and the occupation of trap states in a MAPbBr$_3$ device at electric biases of 100 V a)-b), 20 V c)-d), and 1 V e-f). Figures 5a), c), and e) represent the temporal evolution of free carriers and traps occupation. Flag indicators separate the regions (i)-(iiii) with different

trap E$_3$ further dominates the free carrier trapping, which leads to an exponential decay of the free hole concentration in the region (ii) with the time constant $\tau_{T3}$. At the time $T_R$, the never trapped holes (holes which did not interact with traps) reach the cathode in region **(iii)**. Since a fraction of holes were collected at the electrode, the occupation of all traps decreases after $T_R$. As shown in the bottom panels of Figure 5b, d, and f, due to the presence of several traps participating in trapping-detrapping events and cross retrapping in MAPbBr$_3$, a significant fraction of delayed holes reaches the cathode after $T_R$. Here, the delaying of the charge cloud results in the prolongation of the effective transit time, $T_R^{eff}$, at lower biases in comparison with $T_R$.

Finally, at biases lower than 1 V ($E < 5$ Vcm$^{-1}$) in region **(iiii)**, all traps reach a steady-state condition after processes in the regions (i) and (ii); therefore, the occupation of traps and the free hole concentration do not further change with time. Due to the long $T_R$ ($> \tau_{D3}$), the defect E$_3$ also participates in the hole detrapping in the region (iiii). The steady-state regime in the time region (iiii) is qualitatively similar to the steady-state photoconductivity or solar cell operation regime. Indeed, the continuous illumination utilized in PV devices leads to the redistribution of free charges, and a significant fraction of the photo-generated carriers remains trapped in defects E$_3$.

Note that at low electric field (Figure 5f), the charge cloud does not reach the halfway of the diffusion path in the bulk atthe transit time $T_R$ (corresponding to the never trapped holes) due to the delaying effect of traps. This example explicitly demonstrates the highly detrimental influence of traps on the charge drift in OMHP devices. Besides decreasing the charge collection efficiency, the trapped carriers also contribute to the memory effect and the variation of device transport parameters based on the rate of scanning (up to few ms) and the illumination intensity.

**Delaying effect of traps on the charge transport**
Using the ToF combined with the MC simulation it is demonstrated that the activity of traps leads to a noticeable delay of the charge cloud. The effective mobility, $\mu_h^{eff}$, can qualitatively describe this delaying process. Therefore, the conventional model[55,60] considering a single trap in the material can be modified for the case of multiple traps (see Equations S9-S16 in supplementary materials). If the drift of holes is mainly limited by shallow traps, the effective hole mobility $\mu_h^{eff}$ is given by

$$\mu_h^{eff} = \mu_h \frac{1}{1 + \sum_i \frac{\tau_{Di}}{\tau_{Ti}}}. \tag{3}$$

However, this definition is only valid when all traps reach a saturation point, the condition attainable only at low biases < 1 V (E < 5 Vcm$^{-1}$). Therefore, a new definition is necessary to describe $\mu_h^{eff}$ at higher biases as well.

To modify the effective mobility, $\mu_h^{eff}$, we track the center of the charge cloud, which drifts in a semiconductor with multiple traps. Indeed, the MC simulation allows us to determine the effective transit time, $T_R^{eff}$, of the total charge cloud and to calculate the corresponding effective mobility. The effective mobility can be defined by the following equation,

$$\mu_h^{eff} = \mu_h \frac{T_R}{T_R^{eff}} \tag{4}$$

where the effective transit time, $T_R^{eff}$, describes how the traps delay the drift of free charge cloud. Note that the trapping/detrapping distorts the charge cloud significantly (Figure 5b, d, and e). Thus, the drift mobility ($\mu_h$) cannot be determined by only ToF measurements without considering the delaying effect of traps. The value of $\mu_h$ unaffected by shallow traps can be determined by considering the trapped and detrapped carriers in the MC simulation.

To analyze the influence of each defect level on hole transport and the effective hole mobility $\mu_h^{eff}$, the MC simulation is performed to obtain the dependence of $\mu_h^{eff}$ on the electric field at different MAPbBr$_3$ thicknesses (**Figure 6**). At a high electric field, the electrode collects the holes so rapidly that the effect of traps is negligible. Indeed, Fig. 6 shows that $\mu_h^{eff}$ for all material thicknesses converge to the drift mobility unaffected by traps i.e. 12.4 cm$^2$V$^{-1}$s$^{-1}$ The drift mobility obtained by ToF with MC simulations agrees well with experimentally measured drift mobility of 10-20 cm$^2$V$^{-1}$s$^{-1}$ [20,61] which is not affected by traps.

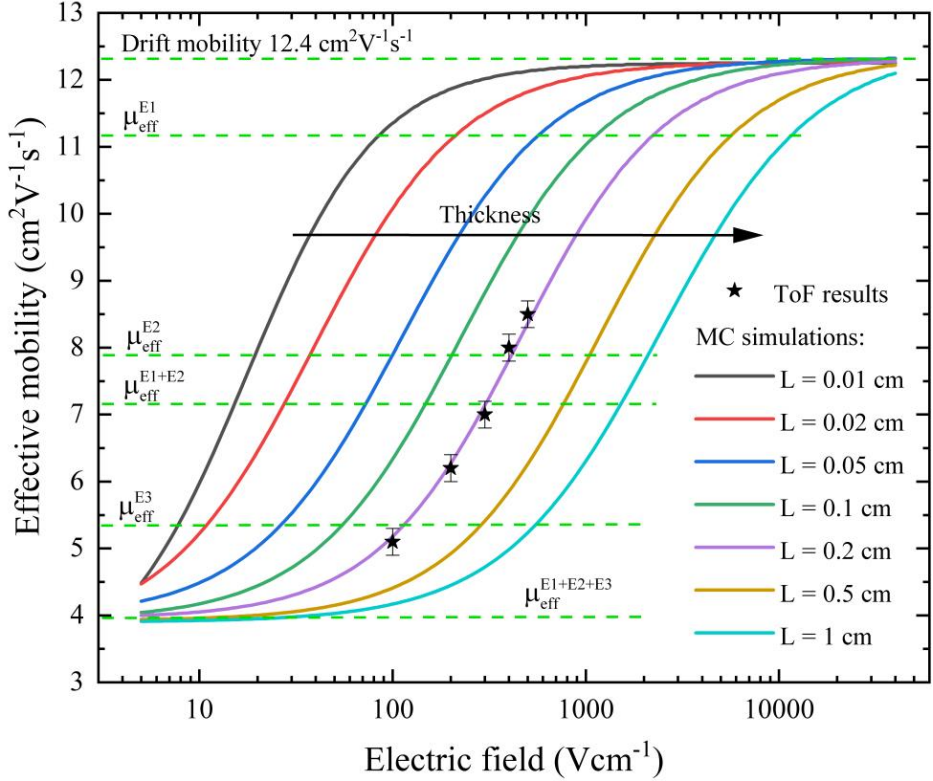

**Figure 6.** The effective hole mobility as a function of the electric field, as determined by the MC simulation for various thicknesses (*L*) of the material. The effective hole mobilities measured by the ToF for *L*=0.2 cm are shown by the stars.

Reducing the electric field leads to stronger interactions between holes and traps; thereby, reducing $\mu_h^{eff}$ as seen in Figure 6. The initial decrease of $\mu_h^{eff}$ to $\mu_{eff}^{E1}$ and further down to $\mu_{eff}^{E2}$ as the electric field is reduced is primarily caused by the traps $E_1$ and $E_2$. Here, $\mu_{eff}^{E1}$ ( $\mu_{eff}^{E2}$) is the effective hole mobility limited by a single trap $E_1$ ($E_2$) under the steady-state condition. $\mu_{eff}^{E1}$ and $\mu_{eff}^{E2}$ are calculated to be 11.0 cm$^2$V$^{-1}$s$^{-1}$ and 7.8 cm$^2$V$^{-1}$s$^{-1}$, respectively, based on Equation 3 and trapping/detraping ratios in Table 1. After $\mu_h^{eff}$ is reduced below $\mu_{eff}^{E2}$ (7.8 cm$^2$V$^{-1}$s$^{-1}$), the further decrease of $\mu_h^{eff}$ down to $\mu_{eff}^{E1+E2}$ (7.2 cm$^2$V$^{-1}$s$^{-1}$) with the decreasing electric field is primarily due to the cross retrapping of detrapped holes involving both traps $E_1$ and $E_2$. Indeed, as demonstrated in Figure 5a, c, d, the trap $E_1$ is the first to reach the steady-state condition and, after $t = \tau_{D1} = 3$ μs, this level actively detrappes holes, which are further trapped and detrapped by the trap $E_2$.

At low electric fields (for example, 300 Vcm$^{-1}$), the $T_R$ of the charge cloud is higher than $\tau_{D3}$; therefore, the trap $E_3$ can effectively participate in the delay of holes cloud, and further decreases the effective hole mobility. The decrease of $\mu_h^{eff}$ below $\mu_{eff}^{E3} = 5.3$ cm$^2$V$^{-1}$s$^{-1}$ (which is the effective hole mobility limited by a single trap $E_3$) is mainly due to the cross re-trapping processes involving all three traps interacting with delayed holes. At sufficiently low electric fields, $\mu_h^{eff}$ saturates at $\mu_{eff}^{E1+E2+E3}$ (4.0 cm$^2$V$^{-1}$s$^{-1}$), which is close to the value obtained from the classical equation (Equation 3) valid for the case in which all three traps reach the steady-state condition. These results clearly demonstrate that the proposed model of effective mobility correctly describes the complicated carrier trapping, detrapping, and re-trapping involving multiple traps during the charge transport process in MAPbBr$_3$ single crystals.

Next, we discuss the thickness dependence of the effective mobility. At the constant electric field, a thicker MAPbBr$_3$ device leads to a longer transit time of the drifting charge cloud and, consequently, a lower $\mu_h^{eff}(E)$ according to Equation S7 (see Figure 6). In thin film devices the $\mu_h^{eff}(E)$ rapidly converges to the drift mobility as the free holes quickly reach the collecting electrode with little interactions with traps. In contrast, a large number of trapping-detrapping events take place in thick samples where the charge cloud interact significantly with traps along its long diffusion path. Therefore, $\mu_h^{eff}$ slowly converges to the drift mobility with increasing electric field in thick samples, and $\mu_h^{eff}$ has a lower absolute value compared to those in thin devices under the same electric field.

With the effect of traps demonstrated in Figure 6, we further discuss the effect of temperature on the effective mobility. In general, phonon scattering dominates the temperature dependence of the drift mobility (following the power-law of $\mu \sim T^{-1.5}$)[62], which, in turn, influences the temperature dependence of the effective mobility. In addition, the temperature dependence of the effective mobility is also affected by the trap activities and can be modeled in the steady-state condition following Equation 3 and Equations S10-S11. Briefly, the charge trapping rate varies slightly with temperature, whereas the detrapping rate is reduced more rapidly with decreasing temperature. Therefore, although lowering temperature reduces phonon scattering, which tends to increase the effective mobility, it also suppresses detrapping of charge carriers from traps; thereby, lowering the effective mobility. These two effects combine to give the temperature dependence of the effective mobility shown in **Figure 7**. Clearly, a strong deviation of the effective mobility from the $T^{-1.5}$[62–64] dependence can be seen as temperature is reduced, especially for T < 250 K, below which the detrapping activity is frozen. Indeed, previous experiments based on time resolved methods (e.g. ToF, transient space-charge-limited currents, etc)[62,64,65] showed deviations of the measured drift mobility from the $T^{-1.5}$ dependence. These experiments, however, did not address the effect of traps on the charge delaying. The present work based on MC simulations provides an explanation to this deviation. Here, we demonstrate a theoretical prediction which agrees with the temperature effect on the effective mobility reported in organic semiconductors[66].

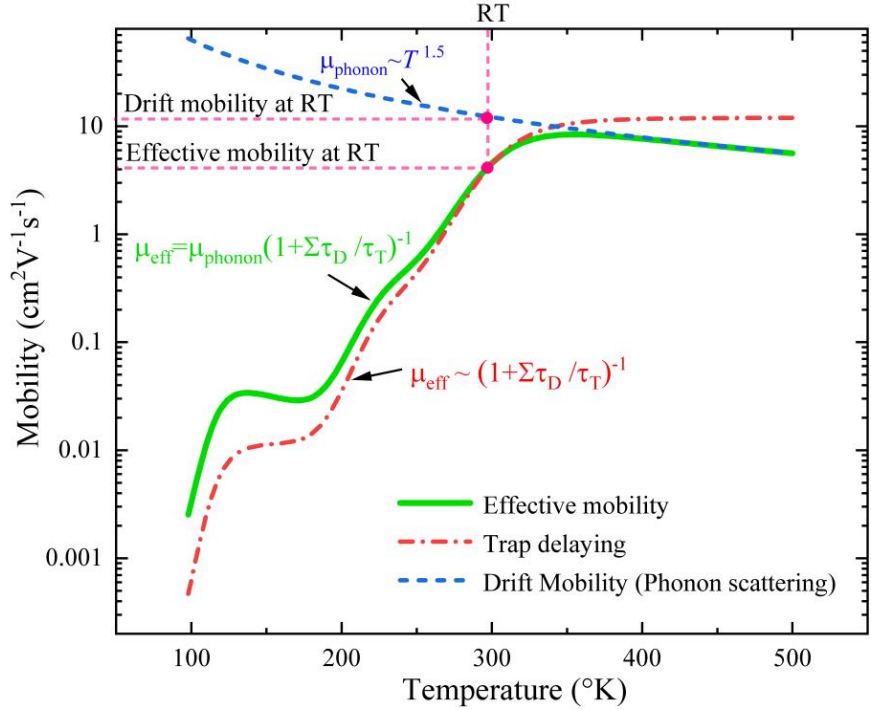

**Figure 7**. The temprature dependence of the effective mobility obtained by theoretical modeling. The observed behavior suggests that traps affect $\mu_{eff}(T)$ dependence in MAPbBr$_3$ single crystals especially at T<250 °K. The experimental drift and effective mobilities (measured at room temperature in steady-state conditions) are also shown by red dots.

**Analysis of drift mobility, lifetime, and mobility-lifetime product ($\mu\tau$) by ToF with MC in MAPbBr$_3$**

Several ToF studies used the inflection (trap free approach[67]) or intersection (dispersive photocurrent[68]) points between transit and tail regions of CWF (see details in Figure S6, supplementary materials) to evaluate the drift mobility in both organic and inorganic semiconductors. The possible errors in the determination of the drift mobility using the above approach is evaluated below. **Figure 8**a compares the drift mobilities calculated by a MC simulation and by the standard methods. The hole mobilities found from inflection ($\mu_{inf}$) and intersection ($\mu_{inter}$) transit times show notable deviations (up to 7.8 cm$^2$V$^{-1}$s$^{-1}$ at 100 Vcm$^{-1}$) from the drift mobility in the MC simulation. The deviation increases at low biases, which agrees with the MC simulation. Note that the MC simulation predicts a larger number of trapping-detrapping events induced by traps and more significant deformation of the hole cloud at lower biases (see Figure 5). Thus, the deformed charge cloud leads to an incorrect treatment of ToF results by simplified approaches, which do not include the effect of traps on the deformation of the charge cloud.

The trapping time of 90 µs from the level $E_3$ obtained in this work is in agreement with our previous calculation of the lifetime of free holes limited by the long-term trapping in MAPbBr$_3$[27]. Such a low trapping time of free holes highlights the advanced transport properties of MAPbBr$_3$ devices. However, the interplay of free charge carriers with traps has a detrimental effect on the performance of OMHP detectors due to the decreased effective mobility and long-term trapping. It is demonstrated that the free carrier dynamics follow a complex non-exponential decay with fast (23 and 24 µs) and slow (90 µs) trapping characters accompanied by the fast (3 and 14 µs) and slow (120 µs) detrapping of carriers, respectively. The release of trapped charge carriers from defects in MAPbBr$_3$ single crystals leads to a significant divergence of lifetime values measured in dynamical (0.3 [69], 1 [43], and 15 µs[29]) and steady-state (> 1 ms[70]) conditions. The fast trapping is typically interpreted as the charge carrier lifetime in time resolved measurements, while the long-term (slow) trapping influences a steady-state lifetime. Thus, the detrapping and the resulting complex carrier decay must be considered for the correct characterization of OMHPs. The significant role of detrapping in the lifetime measurements were also recently discussed by Lang et al[71]. A trapping /detrapping ratio of 3.7 estimated by averaging the values in Table 1 is in agreement with 3.4 in mixed hybrid perovskites reported previously[71]. In addition, Lukosi et al. estimated a trapping time of 26 µs and detrapping time of 15 µs by using a single-trap model to describe the carrier relaxation in MAPbBr$_3$ single crystal detectors[20]. Despite the limitation of the single-trap model[53] these results agree with parameters of the trap $E_2$ in Table 1.

In general, the trapping and detrapping phenomena are scarcely studied in OMHPs. Thus it is useful to compare trap parameters ($\tau_T/\tau_D$) in MAPbBr$_3$ with those in inorganic semiconductors such as GaAs[72] (shallow level 250 ns/40 ns and deep level 150 ns /~3 µs) and CdZnTe (shallow level 13 ns/11 ns and deep level 2 µs /~20 µs). Since the trapping-detrapping times in inorganic semiconductors are much lower, we propose a low capture cross-section of defects in halide perovskites. The low capture cross section can be associated with the strong screening of the charged defects in OMHPs, leading to a low non radiative recombination rate, as previously proposed[73–77]. This effect can be also attributed to the polaronic nature of charge carriers in OMHPs [73,78] or unstable defects configuration after trapping [42,44,79]. In addition to the suppressed capture cross section, the low trapping time in OMHPs can be explained by a low concentration of deep defects with long-term trapping.

Clearly, the interplay between free charge carriers and traps can affect the charge collection properties or mobility lifetime product, µτ. The results of ToF and MC simulations are fitted by the Hecht equation[80] to obtain µτ in MAPbBr$_3$ single crystals, (**Figure 8**b and the inset). Due to the decay of the hole transport by by traps, µτ shows a strong dependence ($10^{-1}$ - $10^{-4}$ cm$^2$V$^{-1}$) on the collection time, which is the time needed to detect high energy absorption events. The detrimental influence of traps can be controlled by eliminating traps, particularly trap $E_3$, or by decreasing their concentrations as shown by MC simulations (Figure 8b). The longer detrapping time of 120 µs from the defect $E_3$ can compete with the transit time of the free charge cloud and the typical charge collection time of classical inorganic semiconductors (typically < 200 µs[81,82]). By considering the collection of free holes in 200 µs, the MC simulation in combination with ToF result in $\mu\tau = 10^{-3}$ cm$^2$V$^{-1}$ in MAPbBr$_3$ devices. This is competitive with the best mobility-life time product in inorganic detectors such as CdZnTe ($10^{-3}$ cm$^2$V$^{-1}$) [58,72], GaAs ($6\times10^{-4}$ cm$^2$V$^{-1}$)[72].

Here, a diffusion coefficient of 0.32 cm$^2$s$^{-1}$ and diffusion length of 54 µm are calculated in MAPbBr$_3$ single crystal devices by considering the mobility and long-term trapping found from our ToF measurements and MC simulations. The results agree well with the previously reported diffusion coefficient, 0.27 cm$^2$s$^{-1}$[65] and diffusion length of 2.6-650 µm[70,83]. A solar cell is usually fabricated using a very thin active layer of ~300-500 nm, thus the charge carriers can be separated by a rather large internal electric field induced by the different work functions of electrodes (~ 1 eV). Consequently, the transit time in a solar cell is on the order of ns, which is much shorter than the trapping times found in this study. Therefore, the demonstrated traps are not as critical in a PV device as in ionizing radiation detectors made from bulky single crystals.

However, in solar cells the high intensity of illumination yields a steady-state hole density of more than $10^{13}$ cm$^{-3}$ (*p=Photon Flux·T$_R$/thickness*). Such a high density of photogenerated carriers could completely fill the traps (holes) up to the position of the quasi-fermi energy $E_F=E_v+0.35$ eV[84]. This result has three crucial implications: (I) the strong carrier trapping induces significant space charge[85,86], which could completely screen the photo-voltage at the density above $10^{14}$ cm$^{-3}$, (II) traps populated by holes could subsequently trap electrons, resulting in non-radiative recombination and reduce $V_{OC}$ in OMHP solar cells. The trapping-detrapping activity decreases the effective mobility, which can consequently lead to instability of photo-voltage (after light is switched on/off), and cause a memory effect (hysteresis)[87]. (III) The typically higher concentration of defects in thin films e.g. by two orders of magnitude[88,89] can lead to a much faster trapping times of defects $E_1$ and $E_2$. The decrease of the trapping time, in turn, can transform the traps with shallow character into fast trapping centers as the detrapping time does not depends on the trap concentration. For example, the trap $E_2$ can increase its trapping time up to ~240 ns in thin films.

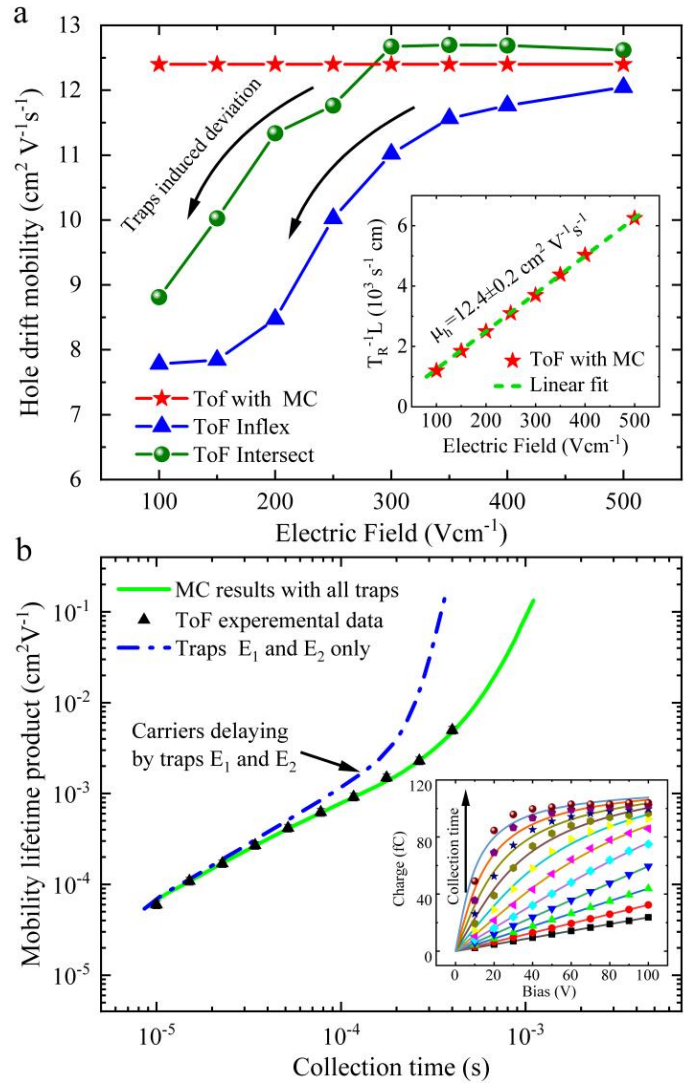

**Figure 8.** a) Holes drift mobility as a function of applied electric field obtained by different approaches using data from Figure 4a and MC simulations. The inset demonstrates the linear fit of the transit time *vs* electric field. b) Mobility-lifetime product as a function of collection time. The Hecht fit is shown in the inset of Figure 8b.

The trap density can be reduced by materials processing such as optimization of growth parameters, doping, or purification. For example, it was shown recently that combining Cs and Rb in quadruple cation (Rb-Cs-FA-MA) perovskite mixtures increases the effective mobility and decreases the trap density, resulting in solar cells with the highest stabilized power efficiency[90].

**Conclusion**

By combining the time of flight current spectroscopy and the Monte Carlo simulation, we have provided novel insights and detailed information on the dynamics of free charge carriers in MAPbBr$_3$ single crystal devices. We have demonstrated strong trapping activities in MAPbBr$_3$ single crystals associated with three defects with trapping times of $\tau_{T1}$ (23 µs), $\tau_{T2}$ (24 µs), and $\tau_{T3}$ (90 µs). While the traps showed fast detrapping times of 3, 14, and 120 µs from the three discovered traps, respectively. Our results reveal that traps have a significant impact on the free carrier transport and the collection efficiency in MAPbBr$_3$ perovskite devices; the traps E$_1$ and E$_2$ act as shallow levels with short-term trapping-detrapping characters while the trap E$_3$ acts as a deep level causing long-term trapping of free holes.

To describe the delaying effect of traps on the charge transport properties of MAPbBr$_3$ single crystals, a new model was proposed for the effective mobility, $\mu_h^{eff}$. It was shown that, by decreasing the electric field, $\mu_h^{eff}$ decreases from the drift hole mobility (12.4 cm$^2$V$^{-1}$s$^{-1}$) to the effective mobility calculated from the classical theory (4 cm$^2$V$^{-1}$s$^{-1}$). Detailed MC simulations show that, besides the trapping-detrapping events from defects, the retrapping of the delayed holes play a significant role in the delaying of the total hole cloud. In addition, the material thickness and temperature are found to have a substantial impact on $\mu_h^{eff}$ and consequently on the free charge collection efficiency in MAPbBr$_3$ devices. Our results demonstrate that the role of traps should be carefully considered in the study of transport properties of OMHP devices, such as the assessment of drift mobility, lifetime, and mobility lifetime product

## Methods

**Time of Flight spectroscopy**

Time-of-Flight (ToF) method is based on measurement of the current response in a planar semiconductor device with an external stimulation (such as an alpha particle, electron, short light pulse, etc.). In this experimental setup, the external stimulus is an above bandgap (450 nm) laser pulse (1 µs) with an intensity power of 0.03 µW. The attenuation depth of the optical photons is estimated to be 1 µm. The optical pulse generates electron-hole pairs near the illuminated electrode. By applying an electric field, the electron-hole pairs are separated. Free holes drift towards the cathode and generate a current according to the Shockley-Ramo theorem. The anode immediately collects free electrons therefore only free holes drifting in the material generates Current waveforms (CWFs) signal. In these measurements, the CWFs is recorded by a synchronous triggering derived from the laser pulse. This set up results a much better signal to noise ratio as compare to un-triggered sources like alpha particles. Often the enhanced continuous DC biasing (tens of V) in an OMHP device results in dynamic degradation of the sample, which makes the reliable record of CWF impossible. To overcome this detrimental effect, we apply an synchronized pulsed biasing. A light source with a photon energy of 2.8 eV (450 nm) is used to generate free carriers at the anode. The above-bandgap light pulse is preferably absorbed in less than 1 µm thick layer below the contact electrode. A positive bias, U, is applied between the two electrodes to collect free charge carriers. The current signal is detected using an oscilloscope synchronized with laser and voltage pulses. Using such pulse photo-excitation allows us an accumulation and averaging of multiple current waveforms resulting in a high signal to noise ratio. The additional description can be found in supplementary materials (Figures S4-S5).

**Monte Carlo (MC) simulation**

We use MC calculation to simulate charge dynamics in MAPbBr$_3$ single crystal device. We developed 1D MC with the total number of particles, $N = 10^5$. The initial position x of the MC particle is generated according to Lambert-Beed law for light absorption. Our MC simulation also includes the diffusion of the carriers in addition to simulated drift process between two metal contacts. Each MC simulation step changes the state of the MC particle using random numbers according to probability given by trapping and detrapping time of particular trap level. We found parameters (trapping/detrapping time) of each trap in the band gap by fitting experimental ToF results with MC simulation and least square regression analysis. Current waveforms are calculated using Shockley-Ramo theorem. The detailed description of the method is given in supplementary materials (Figures S1-S2 and Equations S1-S6).

The effective mobility is a useful approximation of free charged carrier cloud movement. When free carriers drift in the material and interact with traps, the center of the charge cloud moves with an effective mobility rather than with microscopic one. This approximation relies on central limit theorem which states that when carriers undergo many trapping and detrapping events, a new equilibrium between traps and conduction band is established The accuracy of the effective mobility depends on the number of trapping-detrapping events. If there is more than one trap level, the thermalization of traps starts with smallest detrapping time, after that the trap with larger detrapping time is thermalized and so on until all traps are thermalized. To reliably apply Monte Carlo simulation the number of trapping-detrapping events has to be several hundred in order to use the effective mobility approach. The typical error of effective mobility in this study is about 0.3%.

**Sample Preparation: MAPbBr$_3$ single crystal growth**

Single crystal of MAPbBr$_3$ is grown from ultra-purity precursors. The sample geometry is $5 \times 5 \times 2$ mm$^3$. Several samples showed the same ToF and MC simulation results, thus the results of one sample are presented in this work. Detail of single crystal growth can be found in our previous work[19].

**Acknowledgements**

A. M., J. P., P.P, M. B., E. B., and R. G., thank the Institute of Physics of Charles University for providing necessary facilities to conduct this research. A. M., J. P., P.P, M. B., E. B., and R. G acknowledge financial support from the Grant Agency of the Czech Republic, Grant No. P102/19/11920S and the Grant Agency of Charles University, project No. 1234119. M.A. and B.D. acknowledge financial support from U.S. Department of Homeland Security (grant # 2016-DN-077-ARI01). M. -H. D. is supported by the U.S. Department of Energy, Office of Science, Basic Energy Sciences, Materials Sciences and Engineering Division.